\documentclass[journal]{IEEEtran}
\usepackage{algorithmicx}
\usepackage{algorithm,algpseudocode}

\usepackage{amsmath,graphicx}
\usepackage{float}
\usepackage{verbatim}
\usepackage{mathrsfs}
\usepackage{pictexwd,dcpic}
\usepackage{cite}
\usepackage{amsthm}
\usepackage{epstopdf}
\usepackage{epsfig,psfrag}
\usepackage{amsfonts,amssymb}
\usepackage{color}

\newtheorem{theorem}{Theorem}

\newtheorem{defn}{Definition}[section]
\newtheorem{myprop}{Proposition}
\newtheorem{myremark}{Remark}

\usepackage{lipsum}
\usepackage{verbatim}
\usepackage{mathrsfs}
\usepackage{enumitem}
\usepackage{float}
\usepackage{tikz}
\usepackage{pgfplots}

\begin{document}
\title{Graphon Filters: Graph Signal Processing in the Limit}

\author{Matthew~W.~Morency,~\IEEEmembership{Student Member,~IEEE} and Geert Leus,~\IEEEmembership{Fellow,~IEEE}}

\maketitle

\section{Abstract}
 
Graph signal processing is an emerging field which aims to model processes that exist on the nodes of a network and are explained through diffusion over this structure. Graph signal processing works have heretofore assumed knowledge of the graph shift operator. Our approach is to investigate the question of graph filtering on a graph about which we only know a model. To do this we leverage the theory of graphons proposed by L. Lovasz and B. Szegedy. We make three key contributions to the emerging field of graph signal processing. We show first that filters defined over the scaled adjacency matrix of a random graph drawn from a graphon converge to filters defined over the Fredholm integral operator with the graphon as its kernel. Second, leveraging classical findings from the theory of the numerical solution of Fredholm integral equations, we define the Fourier-Galerkin shift operator. Lastly, using the Fourier-Galerkin shift operator, we derive a graph filter design algorithm which only depends on the graphon, and thus depends only on the probabilistic structure of the graph instead of the particular graph itself. The derived graphon filtering algorithm is verified through simulations on a variety of random graph models.

\section{Introduction}
\label{sec:intro}
Graph signal processing is an emerging topic which endeavours to explain the evolution of signals supported on an irregular domain, modeled as a graph \cite{Vandergheynst}-\cite{Coutino1}. The interconnections between nodes of the graph in some way capture interdependencies of data supported on the nodes. This information is encoded in graph shift operators, which also capture operations that are performed on the node-supported data. For example, the action of the scaled adjacency matrix on the data supported on the nodes of the graph leads to a weighted sum of the values on the neighbours of every node.

As an irregular domain analogue of classical signal processing, one of the fundamental operations of graph signal processing is that of filtering. The simple graph filter is a polynomial in the graph shift operator, the coefficients of which are chosen to satisfy as closely as possible a desired filter response. The graph frequencies themselves are taken to be the eigenvalues of the graph shift operator. Both finite impulse response (FIR) \cite{ASandryJMoura1} and infinite impulse response (IIR)\cite{EIsufi1} filters have been investigated and designed. The simple graph filter imposes on the network process a uniform rule: each delay is to be multiplied by the same coefficient by every node in the network. More agile and general filters have been designed whereby each node is able to choose a different coefficient for all of its neighbours (node-varying filters) \cite{Segarra}, or each node is able to scale the signal it receives from each of its neighbours independently (edge-varying filters) \cite{Coutino1}.

Disregarding for the moment universal graph filter design \cite{Vandergheynst1}, each of these strategies relies on knowledge of the graph shift operator in order to design the graph filter. In this paper, we investigate the line of questioning raised by Laszlo Lovasz in \cite{LL}. Specifically, how shall we design graph filters when the graph is so large that we cannot store or evaluate the entire graph and its diffusion, but only statistically probe the graph and derive models which in some way capture the graph's structure. To this end,  \cite{LL} and references therein developed a probabilistic and function theoretic framework known as graphons. The word \textit{graphon} itself is a portmanteau of graph and function. These are two-dimensional kernel functions supported on $[0,1]\times[0,1]$. Random graphs can be drawn from a \textit{graphon} by drawing two samples uniformly distributed over $[0,1]$ for each pair of nodes $x_i, x_j,\ \forall i,j$ and interpreting the two samples as coordinates on the \textit{graphon}. The value of the \textit{graphon} at these coordinates is then the parameter $p$ for a Bernoulli trial which decides the connection of the nodes. Random graphs drawn in such a way are structurally ``related'' to the graphon through graph homomorphism densities \cite{LL}, and the spectra of the scaled adjacency matrices associated to such random graphs can be shown to converge to integral operators with the \textit{graphon} as a kernel function \cite{BS}. 

Leveraging these properties of \textit{graphons}, we make three key contributions. The first is to show that a graph filter defined with the scaled adjacency matrix of a kernel-based random graph as its shift operator converges to the graphon filter defined by the same kernel. The major difficulty in performing this comparison is bridging the gap between the graph filter, which is a finite dimensional linear operator, and the graphon filter which is an infinite dimensional linear operator. To accomplish this task, we ``lift'' the graph filter into the infinite dimensional vector space as a step function defined over $[0,1]$. As a corollary of this finding, we draw a parallel to the numerical solution of Fredholm integral equations. Using projections of the kernel onto different sets of orthonormal bases we can calculate the output of the graph filter with a sparse shift operator, as opposed to a dense one, and thus reduce the cost of computation to $\mathcal{O}(N \mathrm{log}(N))$ where $N$ in this case is the size of the basis chosen to represent the Fredholm integral equation. Finally, using the Fourier-Galerkin shift operator, we derive the optimal order-k graphon filtering algorithm.

To differentiate our approach from other graph filtering approaches, the graphon filtering algorithm is agnostic with respect to the actual graph. Although similar to the universal graph filtering approach, our approach differs from it in a fundamental sense. Graph filtering operates on the graph ``frequencies'' and the output of the graph filter is thus some projection onto the scaled eigenvectors of the graph. Thus, while the filter design may be universal with respect to graph frequencies, the filter design is not universal in terms of its input-output relationship. In contrast, the graphon filter is designed precisely based on the input-output relationship within a standardized basis for all classes of graphs. Thus if the filter is reachable for two given classes of graphs, their outputs will be similar with the same inputs. This is more analogous to filtering in classical signal processing. Our algorithm allows the user to both design a distributed graph filter, and also make predictions about the signal value at nodes in an arbitrary graph from the class for which the filter was designed. As such, it presents potentially great computational benefits in the domain of extremely large graphs, as the complexity of the algorithm depends on the graph complexity, which we call its ``frequency content,'' and not its size.

The structure of this paper is as follows. Section \ref{sec:Background} introduces graph signal processing preliminaries. In section \ref{sec:random graphs} the basics of kernel based random graphs are introduced. In section \ref{sec:HSop} we introduce some basic properties of Hilbert-Schmidt operators of which Fredholm integral operators with graphons as their kernels are a subset, and define graphon filters as the counterpart to graph filters. In section \ref{sec:exp_methods}, we introduce the theory of sparse approximations of Fredholm integral equations via expansion methods. In section \ref{sec:fredholm} we investigate the convergence of graph filters to their graphon counterparts. In sections \ref{sec:FG} and \ref{sec:GFD} we define the Fourier-Galerkin shift operator and derive the graphon filtering algorithm. Section \ref{sec:sims} demonstrates the findings of the paper in simulations with several random graph models, which is followed, finally, by conclusions.

In the following, lower-case letters are scalar variables, upper-case letters are scalar constants, bold lower-case letters denote vectors, bold upper-case letters denote matrices, and caligraphic letters denote special mathematical objects such as distributions, sets, graphs, or functions, and double bold letters correspond to a field, e.g. $\mathbb{C}$ representing the complex numbers, or $\mathbb{K}$ representing a general, unspecified (infinite) field.

\section{Networks, Action, and Shift Operators}
\label{sec:Background}

A graph $\mathcal{G}(\mathcal{V},\mathcal{E})$ is a double defined on a set of nodes $\mathcal{V} = \{v_1,\cdots,v_N \}$, and the set of connections between the nodes $\mathcal{E}$. The set of neighbours to a given node is denoted as $\mathcal{E}_j$. A graph signal $\mathbf x \in \mathbb{R}^N$ is one that is supported on the vertices of an undirected graph $\mathcal{G}(\mathcal{V},\mathcal{E})$. The graph shift operator $\mathbf S$ in some way captures the structure of the graph. For example, $\mathbf S = \mathbf A$, the adjacency matrix where $\mathbf A_{i,j} = \mathbf A_{j,i} = 1$ if $i,j\ \in \mathcal{E}$, and $0$ otherwise. There are many other possible shift operators including the graph Laplacian, or weighted versions of the adjacency and Laplacian. Note that since the $\mathcal G$ under consideration is undirected, the graph shift operator $\mathbf S$ is symmetric, and thus has real eigenvalues and a complete set of orthonormal eigenvectors. In graph signal processing, it is assumed that the network structure in some way captures the evolution of the process supported on the graph nodes. However, it is not just the graph connections which influence the evolution of the graph signal, but also the action that the nodes take on the graph signal at each diffusion step. Any action that can be implemented by a linear operator can be represented as

\begin{align}
[\mathbf x]_{j,t} &= a [\mathbf x]_{j,t-1} + \sum_{i \in \mathcal{E}_j} b_{i,j} [\mathbf x]_{i,t-1} \nonumber
\end{align}
where $b_{i,j}$ is the edge-weight connecting the $i$-th and $j$-th nodes in the graph, $a$ is whatever weight the node gives to its own observation at ``time'' $t-1$, and $[\mathbf x]_{i,t}$ is the $i$-th element of $\mathbf x$ at ``time'' $t$. Time here meaning discrete ordered diffusions through the graph shift operator: in other words ``graph time.'' The signal $\mathbf{x}_{j,t}$ depends on information existing on the nodes of the graph at time $t-1$. For example, in the case of the scaled adjacency matrix the operation each node performs on the signal is

\begin{align}
[\mathbf x]_{j,t} &= \frac{1}{N}\sum_{i \in \mathcal{E}_j} [\mathbf x]_{i,t-1} \nonumber
\end{align}
In this paper, we consider only the scaled adjacency matrix $\mathbf S = \frac{1}{N} \mathbf A$ as the graph shift operator.

The Graph Fourier transform (GFT) of a graph signal $\mathbf x$ supported on a graph $\mathcal{G}(\mathcal{V},\mathcal{E})$ is defined as

\begin{align}
\hat{\mathbf x} = \mathbf U^{T} \mathbf x \nonumber
\end{align}
where $\mathbf S = \mathbf U \mathbf D \mathbf U^T$ is the eigenvalue decomposition of the graph shift operator. The eigenvalues of the shift operator are taken to be the ``graph frequencies.'' Similarly, the inverse graph Fourier transform is defined as $\mathbf x = \mathbf U \hat{\mathbf x}$. Though two networks may have the exact same connections, depending on the actions performed at each node, they may have very different eigen-decompositions, and thus different modes and frequencies. Thus they will also have different frequency responses and different filtering operations altogether for the same filter coefficients.

Having defined the GFT for a fixed network and a given node action, the task of graph filtering can be introduced. The GFT is the expression of the signal supported on the network nodes in the modes of the graph. The filtering operation is then on the eigenvalues of the shift operator. Specifically

\begin{align}
\mathbf{y} = \mathbf U \mathbf D' \mathbf U^T \mathbf x \label{eq:abst_filt}
\end{align}
where $\mathbf D'$ is the matrix with the desired frequency coefficients along the diagonal. For example, if we wanted to filter out the $i$-th graph frequency, we would set $[\mathbf D']_{i,i} = 0$. As clear as this definition of a GFT is, it is abstract. It remains unclear how to actually implement the filtering operation on the network through action of the nodes. By eliminating certain modes, the operator in \eqref{eq:abst_filt} will likely develop non-zero entries in positions corresponding to non-existent connections in the graph. This indicates that the ``ideal'' graph filtering operation may not be realizable.

To filter the graph frequencies, a polynomial approach is thus adopted. The graph signal is diffused over the network by applying the shift operator $K$ times, weighted by a coefficient and summed to produce the filtered output. In view of the definition of the ideal graph filter, the design objective for polynomial graph filters is to achieve a desired filter response in the graph frequencies. Specifically, certain graph frequencies are to be attenuated as much as possible, while holding others constant. With sufficiently many taps, any response can be modeled by

\begin{align}
\mathbf H &= \sum_{k=0}^{K-1} h_k \mathbf{S}^k \label{eq:graph_poly_filt}.
\end{align}
Since the filter is expressed through the diffusion over the graph through the action of the nodes, any such filter is, by definition, reachable. The filter in \eqref{eq:graph_poly_filt} has the implicit constraint that every node is constrained to use the same filter coefficient as every other node at each delay. We call such graph filters ``simple.'' This constraint can be relaxed to produce node-varying and even edge-varying filters. In this paper, we consider only simple filters.

\section{Random Graph Models and Graphons}
\label{sec:random graphs}

\subsection{Kernel-Based Models}

Our motivation is to conduct signal processing tasks on large graphs about whose structure we know little. This could be the case in the real world where the graphs are either too large, or too time-varying to have a single reliable representation. Such networks are typically not well-modeled by the classical Erd\"{o}s-Reny\'i random graphs \cite{AB},\cite{LL}. Real-world networks typically have heavy-tailed degree-distributions, exhibit ``small-world'' phenomena, they tend to be clustered, and have neighborhood density higher than the average edge density \cite{AB},\cite{BCLTSSV},\cite{LL}. This has given rise to new, more general random graph models, one of which are Kernel-based models \cite{BJR}. These are characterized by the use of a symmetric kernel function to generate probabilities controlling the formation of edges between nodes.

A kernel-based random graph is a triple $\mathcal G(N,\mathcal{W},\mu)$ where $N$ is the number of vertices in the graph, $\mathcal{W} : [0,1]^2 \to [0,1]$ is a symmetric measurable function, and $\mu$ is a random variable defined on $[0,1]$. To generate a graph from this triple, for each pair of vertices $v_i$ and $v_j$ a sample is drawn from $\mu$. Then each realization of $\mu$ is treated as a coordinate and mapped to a probability by the symmetric function $\mathcal{W}$. This probability is then used to perform a Bernoulli trial to establish if the two vertices are connected, in a manner analogous to the formation of an Erd\"{o}s-Reny\'i graph. The symmetric function $\mathcal{W}$ is called a \textit{graphon} by Lovasz and Szegedy \cite{LL}. In this paper, we assume that $\mu$ is uniform, i.e. $\mu$ is evenly distributed over $[0,1]$.

Conversely, the adjacency matrix of a graph $\mathcal{G}$ induces a graphon in the following way. Consider a weighted graph with vertex weights $\alpha_n$, where $\sum_n \alpha_n = 1$, and edge weights $0 \leq \beta_{i,j} \leq 1$ . Then, to construct a symmetric function, for each element of the adjacency matrix place a square of area $\alpha_i \times \alpha_j$ in the corresponding position of $[0,1]^2$, in which the symmetric function will be the constant edge weight $\beta_{i,j}$. This graphon is denoted $\mathcal{W}_{\mathcal{G}}$. Here we assume $\alpha_i = \frac{1}{N}$, $\beta_{i,j} = 1$. Such a graphon is referred to as an \textit{empirical graphon}.

\subsection{Graph Sequences and Convergence}

Consider a sequence of random graphs $(\mathcal{G}_n)$ with $n \to \infty$ where each $\mathcal G_n$ is drawn from a kernel model $\mathcal G(n,\mathcal{W},\mu)$, for $n \to \infty$. A natural question is whether this sequence converges to any particular object. If so, then, while the graph in question is still random, its emergent properties would remain deterministic. The graph properties that we consider in this paper are homomorphism densities. The homomorphism density of a simple graph $\mathcal{F}$ with $K$ nodes in a given graph $\mathcal{G}$ with $N$ nodes is defined as

\begin{align}
	t(\mathcal{F},\mathcal{G}) &= \frac{hom(\mathcal{F},\mathcal{G})}{hom(\mathcal{F},\mathcal{K}_N)} \nonumber \\
	&= \frac{hom(\mathcal{F},\mathcal{G})}{N^K}
\end{align}
where $hom(\mathcal{F},\mathcal{G})$ is the number of adjacency preserving maps (homomorphisms) $\mathcal{V}(\mathcal{G}) \to \mathcal{V}(\mathcal{F})$, and $\mathcal{K}_N$ is the complete graph on $N$ nodes, $\mathcal{K}_{N}$. Roughly speaking $t(\mathcal{F},\mathcal{G})$ is a ratio between the number of copies of $\mathcal F$ in $\mathcal G$ and the number of copies of $\mathcal F$ in the complete graph on $N$ nodes. The graph homomorphism densities of different graphs $\mathcal{F}$ in $\mathcal{G}$ provide structural information about the graph. For example, if $\mathcal{F} = \mathcal{K}_3$ then $t(\mathcal{F},\mathcal{G})$ is the triangle density of $\mathcal{G}$. A sequence of graphs $(\mathcal{G}_n)$ is said to be $\textit{convergent}$ if $t(\mathcal{F},\mathcal{G}_n)$ converges for any simple graph $\mathcal{F}$ \cite{BCLSV}.

We can further use the symmetric function $\mathcal{W}$ to define a limit object of the graph sequence $(\mathcal{G}_n)$. Let $\mathcal{W}$ be a bounded symmetric function $\mathcal{W} : [0,1]^2 \to [0,1]$, and $\mathcal{F}$ be a simple graph with $K$ nodes. Then the graph homomorphism limit of $\mathcal{F}$ in $\mathcal W$ is defined as
\begin{align}
	t(\mathcal{F},\mathcal{W}) &= \int_{[0,1]^K} \prod_{(i,j)\in \mathcal{E}(\mathcal{F})} \mathcal{W}(x_i,x_j) \mathrm{d}\mathbf x \label{eq:hom_dens}
\end{align}
The following two theorems, proven in \cite{LovaszSzegedy1}, form the fundamental link between convergent graph sequences and graphons, and hence the link between observed graphs and explanatory graphon models.

\begin{theorem}
	For every convergent graph sequence $(\mathcal{G}_n)$ there exists a bounded measurable symmetric function $\mathcal{W} : [0,1]^2 \to [0,1]$ such that $\lim\limits_{n\to \infty} t(\mathcal{F},\mathcal{G}_n) = t(F,\mathcal{W})$
	\label{thm:conv1}
\end{theorem}

\begin{theorem}
	The graph sequence, $(\mathcal{G}_n)$, where $\mathcal{G}_n$ is drawn from $\mathcal{G}(n,\mathcal{W},\mu)$, is convergent with probability 1, and its limit object is the function $\mathcal{W}$. \label{thm:conv2}
\end{theorem}

Thus the graphon itself can be used to study the properties of the graph sequence. As will be observed in the next section, these are not limited to structural properties of the graph, as we can also use the graphon to calculate the spectra of the graphs in $(\mathcal{G}_n)$ as $n \to \infty$.

\section{Fredholm Integral Equations and Graphon Filters}
\label{sec:HSop}

We begin by defining a Hilbert-Schmidt kernel function.

\begin{defn}
	Let $X$ and $Y$ be intervals in $\mathbb{R}$ and $\mathcal{W} : X \times Y \to \mathbb{R}$. If $\int_{X} \int_{Y} |\mathcal{W}(x,y)|^2 \mathrm{d}x\ \mathrm{d}y < \infty$ then $\mathcal W$ is a Hilbert-Schmidt kernel function.
\end{defn}

Graphons, as defined in the previous section are clearly Hilbert-Schmidt kernel functions, being both bounded and measurable. Hilbert-Schmidt kernel functions induce bounded integral operators on the space of square integrable functions $L^2(Y) \to L^2(X)$. In the case of graphons, they induce an integral operator  $T : L^2[0,1] \to L^2[0,1]$

\begin{align}
	T f(x_1) &:= \int_{0}^{1} \mathcal{W}(x_1,x_2) f(x_2) \mathrm{d}x_2. \label{eq:HS_op}
\end{align}

Equations of the form 

\begin{align} \label{eq:fredholm}
g(x) = Tf(x)
\end{align}
are known as Fredholm integral equations of the first kind \cite{Haase}.

Since the graphons are required to be symmetric, the operators they induce are self-adjoint. This, combined with the fact that all Hilbert-Schmidt operators are compact \cite{Haase} implies that the spectrum of the operator \eqref{eq:HS_op} consists of a finite number of real-valued eigenvalues. The eigenvalues of \eqref{eq:HS_op} and corresponding eigenfunctions can be found by solving the resolvent equation

\begin{align}
	\lambda f(x_1) &= \int_{0}^{1} \mathcal{W}(x_1,x_2) f(x_2) \mathrm{d}x_2 \label{eq:HS_resolv}
\end{align}
for $f(x)$ and $\lambda$, given the kernel $\mathcal{W}$. Since we know from Theorems \ref{thm:conv1}-\ref{thm:conv2} that graph sequences converge to their limit object, and that every graph sequence has a limit object, knowledge of the spectrum of \eqref{eq:HS_op} provides knowledge of the spectrum of the graphs in $(\mathcal{G}_n)$, particularly when the number of vertices becomes large. Indeed, it is not difficult to see from the empirical graphons $\mathcal{W}_{\mathcal{G}_n}$ that the spectrum of $\mathcal{W}$ is the spectrum of the scaled adjacency matrix corresponding to  $\mathcal{G}_n$ as $n \to \infty$. Theorem $11.53$ in \cite{LL} gives a formal statement and proof of this fact. Similarly, the degree matrix is also determined by the graphon, and thus, the spectrum of the graph Laplacian can also be investigated through the resolvent equation. We do not pursue this here, though.

As was shown in section \ref{sec:Background} a distributed simple graph filter $\mathbf H$ can be 
implemented as a polynomial in the graph shift operator $\mathbf S$

\begin{align}
\mathbf H &= h_0 \mathbf I + \sum_{k=1}^{K} h_k \mathbf S^k \label{eq:dist_filter},
\end{align}
also known as a $k$-th order FIR graph filter. The graph filter of order $1$ with $h_0 = 0$ is simply the matrix vector product $\mathbf y = \mathbf S \mathbf x$ which as we've mentioned converges spectrally to the equation $g(x) = T f(x)$. The connection between $g(x)$,$f(x)$, $T$ and $\mathbf S$ is the subject of section \ref{sec:fredholm}. By defining the powers of $T$ through operator composition

\begin{align}
T^2 f(x) = T T f(x) = \int_{0}^{1} \mathcal{W} (x,z) \int_{0}^{1} \mathcal{W}(z,y) f(y) dy\ dz.
\end{align}
we arrive at the definition of a graphon filter. The higher powers of $T$ are derived inductively. We draw attention to the fact that there must be an intermediate dimension added for each power of $T$. A graphon filter is then defined as

\begin{align}
\mathcal{H}f(x) &= h_0 f(x) + \sum_{k = 1}^{K} h_k T^k f(x). \label{eq:graphon_filter}
\end{align}

\section{Expansion Methods for Fredholm Integral Equations}
\label{sec:exp_methods}

Comparing \eqref{eq:fredholm} and $\mathbf y = \mathbf S \mathbf x$ there is one fundamental difference which must be reconciled. The Fredholm integral operator in \eqref{eq:fredholm} operates on functions with a continuous domain $[0,1]$, whereas the equation $\mathbf y = \mathbf S \mathbf x$ is over a finite dimensional vector space. We relate the two via the expansion method. Let $\mathcal{B} = \{b_1,b_2,\cdots \}$ be a complete orthonormal basis for $L^2[0,1]$. The expansion method seeks to approximate the (forward) solution of the Fredholm integral equation, as $g(x) = \sum_{i=1}^{\infty} g_i b_i(x)$. We seek to find the coefficients $g_i$ given the graphon $\mathcal{W}(x,y)$, and the input function $f(x)$. To do this, we use the inner product over $L^2[0,1]$ to project both sides of \eqref{eq:fredholm} onto the basis $\mathcal{B}$, as

\begin{align}
\int_{0}^{1} g(x) b_i(x) \mathrm{d}x = \int_{0}^{1}b_i(x) \int_{0}^{1} \mathcal{W}(x,y) f(y) \mathrm{d}y\ \mathrm{d}x \ \nonumber \\ \mathrm{for}\ i \in 1,2,\dots
\end{align}
Substituting the expansions of $g(x)$ and $f(y)$ allows us to write
\begin{align}
g_i = \int_{0}^{1} b_i(x) \int_{0}^{1} \mathcal{W}(x,y) \sum_{j=1}^{\infty} f_j b_j(x) \mathrm{d}y\ \mathrm{d}x
\end{align}
where the left hand side follows from the orthonormality of $\mathcal B$. Inspecting the right hand side, we see that for each $j$, the integral with respect to $y$ can be factored out. The integral with respect to $y$ thus becomes the projection of $\mathcal{W}(x,y)$ onto $b_j(y)$, leading to $\mathcal{W}_j(x) = \int_{0}^{1} \mathcal{W}(x,y) b_j(y) \mathrm{d}y$.
\begin{align}
g_i = \sum_{j=1}^{\infty} f_j \int_{0}^{1} b_i(x) \mathcal{W}_j(x) \mathrm{d}x \label{eq:exp_meth_deriv}
\end{align}
Since the sum is invariant with respect to $x$, the order of summation and integration can be swapped resulting in \eqref{eq:exp_meth_deriv}. The projections of $\mathcal{W}_j(x)$ onto $\mathcal{B}$ are now no longer orthogonal to the basis functions $b_i(x)$. By inspection, for $i$ and $j$ ranging over $1,2,\dots$, \eqref{eq:exp_meth_deriv} can be written as the following matrix vector equation.

\begin{align}
\mathbf g = \mathbf W \mathbf f \label{eq:op_eq}
\end{align}
where $[\mathbf W]_{i,j} = \int_{0}^{1} b_i(x) \int_{0}^{1} \mathcal{W}(x,y) b_j(y) \mathrm{d}x\ \mathrm{d}y$ and $[\mathbf f]_j = \int_{0}^{1} f(y) b_j(y) \mathrm{d}y$. If $\mathcal{B}$ is complete and orthonormal, $g(x)$ can be approximated without error by \eqref{eq:op_eq}.
We call $\mathbf W$ the operator matrix, and \eqref{eq:op_eq} the operator equation. The operator equation is a linear algebraic equation of countably infinite dimension. Thus, by truncating \eqref{eq:op_eq} we arrive at a finite linear approximation of an integral equation over an infinite domain. This truncated finite approximation we denote as
\begin{align}
\mathbf g_a = \mathbf W_a \mathbf f_a. \label{eq:op_eq_trunc}
\end{align}
As $\mathbf g_a$ is a representation of the solution of the Fredholm integral equation in the basis $\mathcal B$, the approximate solution to the integral equation can then be resampled with as many points as desired as
\begin{align}
\mathbf y' = \sum_{i=1}^{N} [\mathbf{g}_a]_i \mathbf{b}_i
\end{align}
where $N$ is the number of approximating functions in the expansion, and $\mathbf b_i$ is the basis function $b_i(x)$ uniformly sampled over $[0,1]$. 

To make an analogy to classical signal processing, \eqref{eq:op_eq_trunc} is like the representation of a ``time'' domain signal \eqref{eq:fredholm} in ``frequency'' domain. 

\begin{myremark}
In the above derivation, we have assumed the orthonormality of the set $\mathcal{B}$. Practically, basis functions which are not even orthogonal can be used by introducing a weighting function $w(x)$, as we do in section \ref{sec:FG}. Normality can be enforced by renormalization.
\end{myremark}

\begin{myremark}
In most cases, $\mathbf W_a$ will be a principal sub-matrix of $\mathbf W$. However, this is not necessarily the case. In the next section, the size of the basis determines the basis functions themselves. To remind the reader of this distinction, we mark such bases as $b'_i(x)$.
\end{myremark}

\begin{myremark} The choice of basis affects the matrix $\mathbf W$. Specifically, the size of the matrix and its sparsity depend both on the choice of the basis and the properties of the graphon itself. Smooth graphons will be easily approximated by a few non-zero coefficients with respect to smooth basis functions, while graphons with abrupt changes will require more basis functions and more non-zero coefficients.
\end{myremark}

\section{Empirical Graphons and the Fredholm Equation}
\label{sec:fredholm}
The purpose of this section is to make explicit the connection between graph and graphon filters. We will do this in a series of steps illustrated in the below commutative diagram. 
\begin{center}[t]
\begindc{\commdiag}[1000]
\obj(1,1)[aa]{$\mathbf y = \mathbf S \mathbf x$}
\obj(1,0)[bb]{$g_a(x) = T_e f_a(x)$}
\obj(2,0)[cc]{$T_e f(x)$}
\obj(3,0)[dd]{$T f(x)$}
\obj(3,1)[ee]{$\mathbf g = \mathbf W \mathbf f$}
\mor{aa}{bb}{$\phi$}
\mor{bb}{cc}{1}
\mor{cc}{dd}{2}
\mor{aa}{dd}{}
\mor{dd}{ee}{Sec. \ref{sec:FG}}
\mor{ee}{aa}{$\mathbf y' = \sum_{i=1}^{N} [\mathbf{g}]_i \mathbf{b}_i$}
\enddc
\end{center}
The link $\phi$ bridges the gap between the finite dimensional shift operator and the infinite dimensional graphon Fredholm operator defined by the empirical graphon corresponding to the graph in question, operating on the piecewise step-functions defined on $[0,1]$. The first right arrow then shows that, since the piecewise step-functions are dense in the continuous functions, that there is a stepfunction arbitrarily close to any element of the function $L^2[0,1]$, and therefore the composition of the empirical operator defined on a piece-wise approximation of a function in $L^2[0,1]$ converges to the empirical operator composed on that function. And finally, the last right arrow concerns the probabilistic convergence of the empirical operator to the true graphon operator. The logic is that the shift operator is injectively related to an object which deterministically converges to something which probabilistically converges to the desired result.

This chain concerns the convergence of the scaled adjacency matrix operating on approximations of the function space $L^2[0,1]$ to the true graphon operator defined over the whole space. Graph and graphon filters, however, are comprised of powers of these operators. The convergence of the powers of $\mathbf S$ to the powers of $T$ is demonstrated in the final subsection.

As a general note, while the bases used in section \ref{sec:exp_methods} were arbitrary complete orthonormal bases, in this section specific bases are used to illustrate key points. Thus, for example, $\mathbf W_a$ in this section is a specific instance of $\mathbf W_a$ from the previous section.

\subsection{From finite to infinite dimensions}

To compare a finite vector to a function in $L^2[0,1]$ we must define the map $\phi$. Given a basis $\mathcal B$ and a finite dimensional vector $\mathbf x$ we define
 
\begin{align}
\phi(\mathbf x) := f_a(x) &= \sum_{i=1}^{N}  [\mathbf x]_i b_i(x),\ b_i(x)\in\mathcal{B}. \label{eq:lift}
\end{align}
The map $\phi^{-1}$ then returns the coefficients $[\mathbf x]_i$. The map $\phi$ could also be similarly defined (though in the opposite sense) given a function $f(x)$, however, our concern is analyzing the convergence of the shift operator $\mathbf S$ to a continuous operator $T$.

\begin{myprop}
There exists a basis $\mathcal{B}$ such that $\phi$ is bijective.
\end{myprop}

\begin{proof}
 Let $\mathcal{G}$ be a kernel-based random graph of $N$ nodes drawn from a kernel $\mathcal{W}(x,y)$. We now show that the evaluation of $T_e f_a(x)$ is equivalent to $\mathbf S \mathbf x$. That is the finite vector $\phi^{-1}(g_a(x))$ where $g_a(x) = T_e f_a(x)$ will equal $\mathbf y$. Assume that all the nodes are evenly weighted with node-weight $\alpha = 1/N$, and connections between nodes $i$ and $j$ are weighted by $\beta_{i,j} = 1$. Then, with $x_i$ being the $i$-th strip of width $1/N$ of the interval $[0,1]$ centered at $i \cdot 1/N$, the empirical graphon $\mathcal{W}_e(x_i,y_j) = 1$ if nodes $i$ and $j$ are connected, and $\mathcal{W}_e(x_i,y_j) = 0$ otherwise.

This empirical graphon can be approximated exactly using the procedure described in the previous section, using a basis of $N$ orthonormal functions. Take as the basis $\{b_1, \cdots, b_N \}$ the functions

\begin{equation}
b'_i(x) = 
\begin{cases} 
N & \text{if } x \in [i\cdot\frac{1}{N}, i\cdot\frac{1}{N} + \frac{1}{N}) \\
0       & \mathrm{otherwise}
 \end{cases}
 \label{eq:basis}
\end{equation}

The approximation $\mathbf W$ of the empirical graphon $\mathcal{W}_e(x,y)$ is then the adjacency matrix of the graph with respect to this basis, for any size $N$. To see this, we use \eqref{eq:op_eq} to calculate $\mathbf W_a$,

\begin{align}
[\mathbf W_a]_{i,j} &= \int_{0}^{1} \int_{0}^{1} \mathbf W_e(x,y) b'_i(x) b'_j(y) \mathrm{d}x \mathrm{d}y \nonumber \\
&= 1 \cdot \frac{1}{N^2} \cdot N^2 = 1\  \mathrm{if}\  (i,j)\  \in\  \mathcal{E},\ 0\ \mathrm{o.w.}
\end{align}
which is the adjacency matrix of $\mathcal{G}$, by definition.

Thus the truncated Fredholm integral operator \eqref{eq:op_eq_trunc} can be written as

\begin{align}
g_a(x) &= \int_{0}^{1} \mathcal{W}_e(x,y) f_a(y) \mathrm{d}y \label{eq:fred} \\
g_a(x) &= \sum_{i=1}^{N}\sum_{j\ \in \mathcal{E}_j} \frac{1}{N} [\mathbf x]_j b'_i(x) \nonumber \\
\implies \phi^{-1}(g_a(x)) &= \frac{1}{N} \mathbf W_a \phi^{-1}(f_a(x)) \label{eq:scaled_adjacency} = \mathbf S \mathbf x = \mathbf y.
\end{align}
 This simple derivation leverages the fact that $\mathbf W_a$ is exactly $\mathbf A$ with respect to the basis $\mathcal{B}$. Having noted this, the integral in \eqref{eq:fred} is equivalently represented by the averaging operation in \eqref{eq:scaled_adjacency}, and thus $\mathbf g_a$ is exactly $\mathbf y$. Thus, there is no error introduced by the replacement of the integral in \eqref{eq:op_eq} by the summation (implemented by the matrix product) and division operations in \eqref{eq:scaled_adjacency}. Equivalence of higher powers of $T_e^k$ follow inductively from the observation that $T_e$ maps stepfunctions to stepfunctions.
\end{proof}

\subsection{Right Arrow $1$}

Let $T_e$ be a Fredholm integral operator with the empirical graphon of $\mathcal{G}$ as its kernel, and $T_a$ be the Fredholm operator in \eqref{eq:op_eq_trunc} with the approximation of $\mathcal{W}_e$ as its kernel, using the basis defined in the previous section. Then, in order to show convergence of $T_a f_a(x) = g_a(x)$ to $T_e f(x) = g_e(x)$ we denote the error function as $e_a = g_a(x) - g_e(x)$. Then, it can be seen that

\begin{align}
T_e f(x) - T_a f_a(x) &= g_e(x) - g_a(x) \nonumber \\
\implies T_e (f(x) - f_a(x)) &= g_e(x) - g_a(x) \nonumber \\
\implies \| e_a \| &\leq \| T_e \| \cdot \|f(x) - f_a(x)\|
\end{align}.

The first step is implied by the fact that 

\begin{align}
\mathcal{W}_e(x,y) &= \sum_{i=1}^{N} \sum_{j=1}^{N} [\mathbf W_a]_{i,j} b'_i(x) b'_j(y)
\end{align}
where $\mathbf W_a$ is as defined in the previous subsection, and the second step uses the Cauchy-Schwarz inequality. It is not, in general, true that $T_e$ and $T_a$ as defined at the beginning of this subsection will be equal, it is only through the specific choice of basis that this holds. The norm $\|\cdot\|$ is with respect to the space $L^2[0,1]$.

In this case, the approximation error is entirely due to the quadrature error of the input $f(x)$, and is scaled by the spectral radius ($T$ is self-adjoint) of $T$. We know that this is less than $1$ from \cite{LL}. Thus, to guarantee convergence of the matrix approximation of the empirical graphon Fredholm equation by \eqref{eq:op_eq} as $N \to \infty$, we only need to assume continuity of the input $f(x)$. As a note, the continuous functions are dense in $L^2[0,1]$, and as $N \to \infty$, the set of basis functions \eqref{eq:basis} are a basis for this space.

\begin{figure}[t]
	\begin{center}
		\includegraphics[width=9.3cm,trim=20 225 0 215]{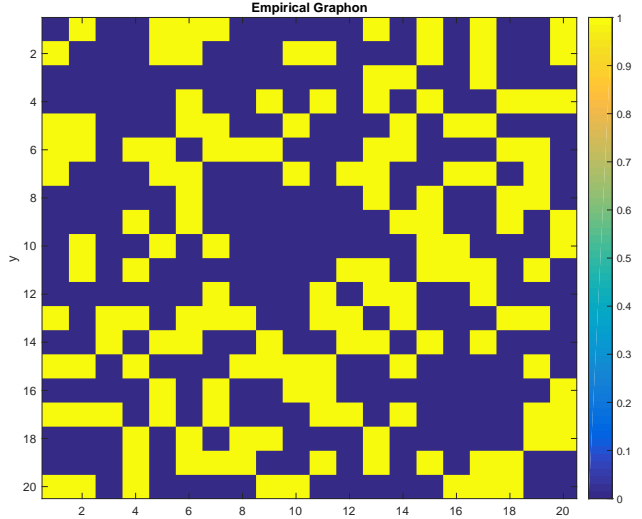}
	\end{center}
	\label{fig:ER09}
	\caption{Empirical graphon of a random graph of 20 nodes generated from the graphon $\mathcal{W}(x,y) = e^{-1/2(x + y)}$}.
\end{figure}

With the previous result in mind, we return to the comparison between \eqref{eq:dist_filter} and \eqref{eq:graphon_filter} with the goal of bounding the total error of the approximation of \eqref{eq:graphon_filter} defined by the empirical graphon by \eqref{eq:dist_filter} where $\mathbf S$ is given by the scaled adjacency matrix.

Define the output of a filter acting on $L^2[0,1]$ as $z_e(x) = (I h_0 + \sum_{k=1}^{K} h_k T_e^k)f(x)$ and the output of a graph filter using the scaled adjacency matrix lifted into $L^2[0,1]$ via the equivalence relationship between \eqref{eq:op_eq_trunc} and \eqref{eq:scaled_adjacency} as $z_a(x) = (h_0 I + \sum_{k=1}^{K} h_k T_a^k)f_a(x)$. We emphasize that $z_a(x)$ is a function in $L^2[0,1]$, but it is equivalently represented as a vector in $\mathbb{R}^N$ as the output of a graph filter defined by the scaled adjacency matrix. Then, taking the norm 

\begin{align}
\|z_e(x) - z_a(x) \| &= \|h_0 I(f_e(x) - f_a(x)) - h_1 T_e(f_e(x) - \nonumber \\ &f_a(x)) + \cdots + T_e^k (f_e(x) - f_a(x))\| \nonumber \\
&\leq \|\sum_{k=0}^{K} h_k T_e^k\| \cdot \|f(x) - f_a(x)\| \nonumber \\
\end{align}

Since the spectral radius of $T_e$ is strictly less than $1$, $\|T_e^k\| < \|T_e\|$ yielding the result

\begin{align}
\|z_e(x) - z_a(x) \| &< \|\sum_{k=0}^{K} h_k\|\cdot \|T_e\| \cdot \|f(x) - f_a(x)\|
\end{align}

As the dimension $N \to \infty$, the approximation $\|f(x) - f_a(x)\|$ tends to zero assuming $f(x)$ is continuous. The quantities $\|\sum_{k=0}^{K} h_k \|$ and $\|T_e\|$ are constants with respect to the limit as $N \to \infty$ and thus the graph filter defined over the scaled adjacency matrix converges to the graphon filter defined over the empirical graphon.

\subsection{Right Arrow $2$}

The previous sections define the sense in which the diffusion of a signal on the nodes of a graph converge to the solution of a Fredholm integral equation. However, the Fredholm integral operator in the previous section has as its kernel the empirical graphon. What remains to be shown is how the output of the empirical Fredholm equation relates to the Fredholm equation associated with the ``true'' graphon.

Let $g_e(x) = \int_{0}^{1} \mathcal{W}_{e}(x,y) f(y) dy$, then the error we wish to quantify is $\|g(x) - g_e(x) \|$.

\begin{align}
\|g(x) - g_e(x)\| &= \|\int_{0}^{1} \mathcal{W}(x,y) f(y) dy - \nonumber \\ &\int_{0}^{1} \mathcal{W}_e(x,y) f(y) dy \| \nonumber \\
&= \|\int_{0}^{1} (\mathcal{W}(x,y) - \mathcal{W}_e(x,y)) f(y) dy \| \nonumber \\
&\leq \| \mathcal{W}(x,y) - \mathcal{W}_e(x,y) \| \cdot \| f(y) \| \nonumber \\ \label{eq:converge}
\end{align}
where the second step follows from Cauchy-Schwarz. 

In \cite{LL}, it is shown that, in general $ \| \mathcal{W}(x,y) - \mathcal{W}_e(x,y) \|_{\square} $ where $\|\cdot\|_{\square} = \sup_{S,T \subset [0,1]} |\int_{S \times T} \mathcal{W}(x,y)\ \mathrm{d}x\ \mathrm{d}y|$, does not converge to $0$ as $N \to \infty$, as the cut norm varies widely with the node labeling of the graph to which the empirical graphon corresponds. The cut metric defined as
\begin{align}
\delta_{\square}(\mathcal{W},\mathcal{W}_e) \triangleq \inf_{\psi} \|\mathcal{W} - \mathcal{W}_e^{\psi}\|_{\square}
\end{align}
where $\psi$ is an invertible, measure preserving map on $[0,1]$, and $\mathcal{W}_e^{\psi} = \mathcal{W}_e(\psi(x),\psi(y))$ can be shown to converge.
  Since $\|\mathcal{W}\|_{\square} \leq \|\mathcal{W}\|$ it is certainly true that $\|\mathcal{W}-\mathcal{W}_e\|$ does not converge, in general. However, it was shown in Corollary 1.1 of \cite{BS} that if $\|\mathcal{W}_e\|$ converges to the spectral norm of $\|\mathcal{W}\|$, and $\|\mathcal{W} - \mathcal{W}_e\|_{\square}\ \to\ 0$, then $\mathcal{W}_e$ also converges in the $L_2$ topology. Therefore, the conditions under which $\|\mathcal{W}(x,y) - \mathcal{W}_e(x,y)\| \to 0$ should be investigated.

In \cite{AvellaMedina}, several rates for $\mathcal{W}_e(x,y)$ to converge to $\mathcal{W}$ are given assuming that the sample $\mu_i$ are ordered, i.e. $\mu_1 \leq \mu_2 \leq \cdots \leq \mu_N$. In the first case, the graphons are assumed to be piece-wise Lipschitz continuous. In the appendix, however, they give a convergence rate for a general graphon, based on Lemma $10.16$ of \cite{LL}. For the purposes of completing the commutative diagram, it is sufficient to note one such rate.

\begin{myprop}
If $\mu_i$ are ordered, then, with probability at least $1-\mathrm{exp}(-\frac{N}{2\mathrm{log}N})$, $\|\mathcal{W}_e(x,y) - \mathcal{W}(x,y)\| \leq \sqrt{176}/\mathrm{log}(N)^{1/4}$.
\end{myprop}

In particular, $\mathrm{lim}_{N \to \infty} \|\mathcal{W}_e(x,y) - \mathcal{W}(x,y)\| \to 0$.

Informally, the sorting of $\mu_i$ introduces a labeling, thus eliminating the need for the measure preserving map $\psi$. Visually, the effect of sorting on $\mu_i$ conveys the point of proposition $2$ quite clearly.

\begin{figure}[b]
	\begin{center}
		\includegraphics[width=9.3cm,trim=20 225 0 215]{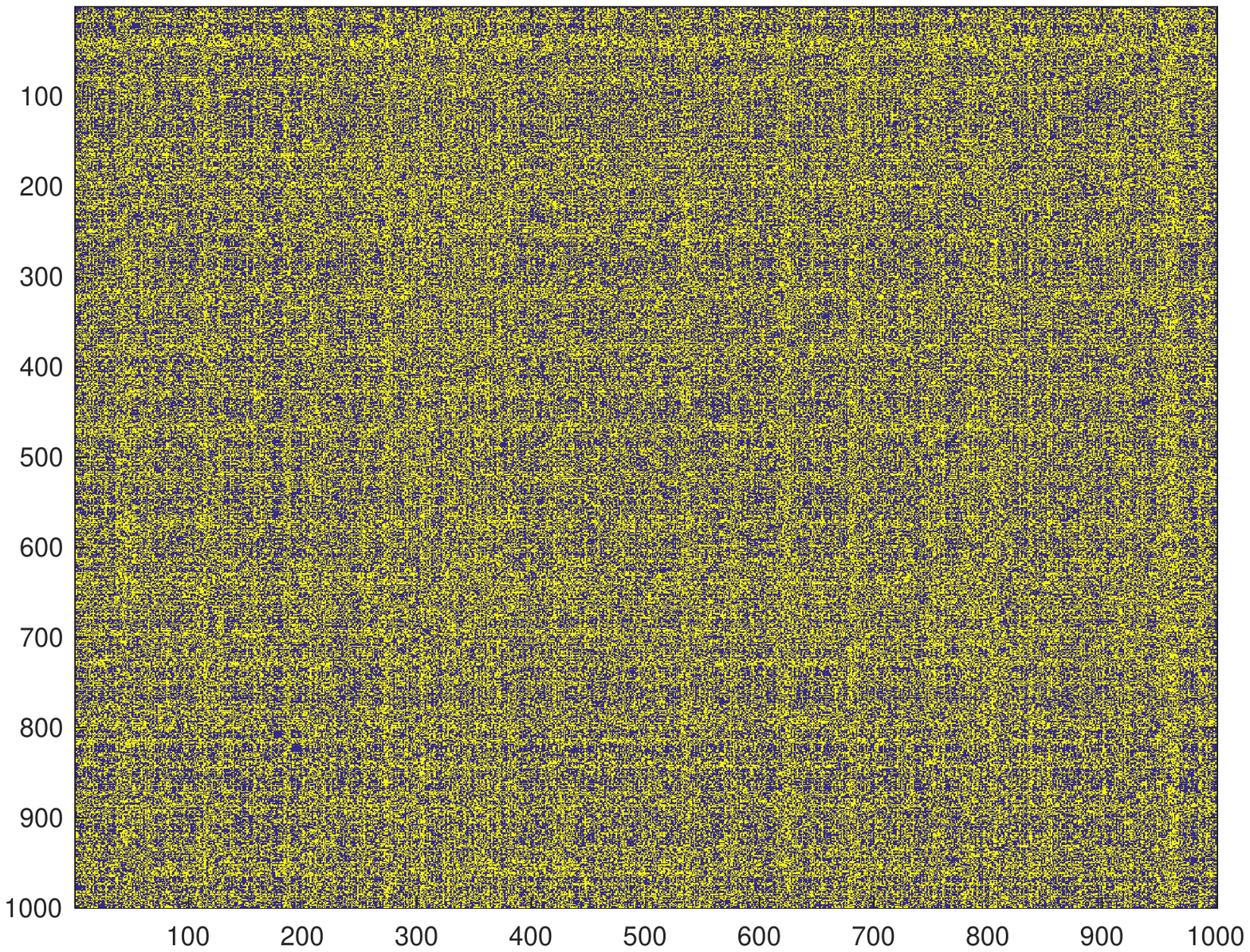}
	\end{center}
	\label{fig:unsorted}
	\caption{Empirical graphon corresponding to unsorted samples of $\mathcal{W}(x,y) = 1/3 + 1/3\cdot \mathrm{sin}(3\cdot \pi x \cdot y)$}.
\end{figure}

\begin{figure}[t]
	\begin{center}
		\includegraphics[width=9.3cm,trim=20 225 0 215]{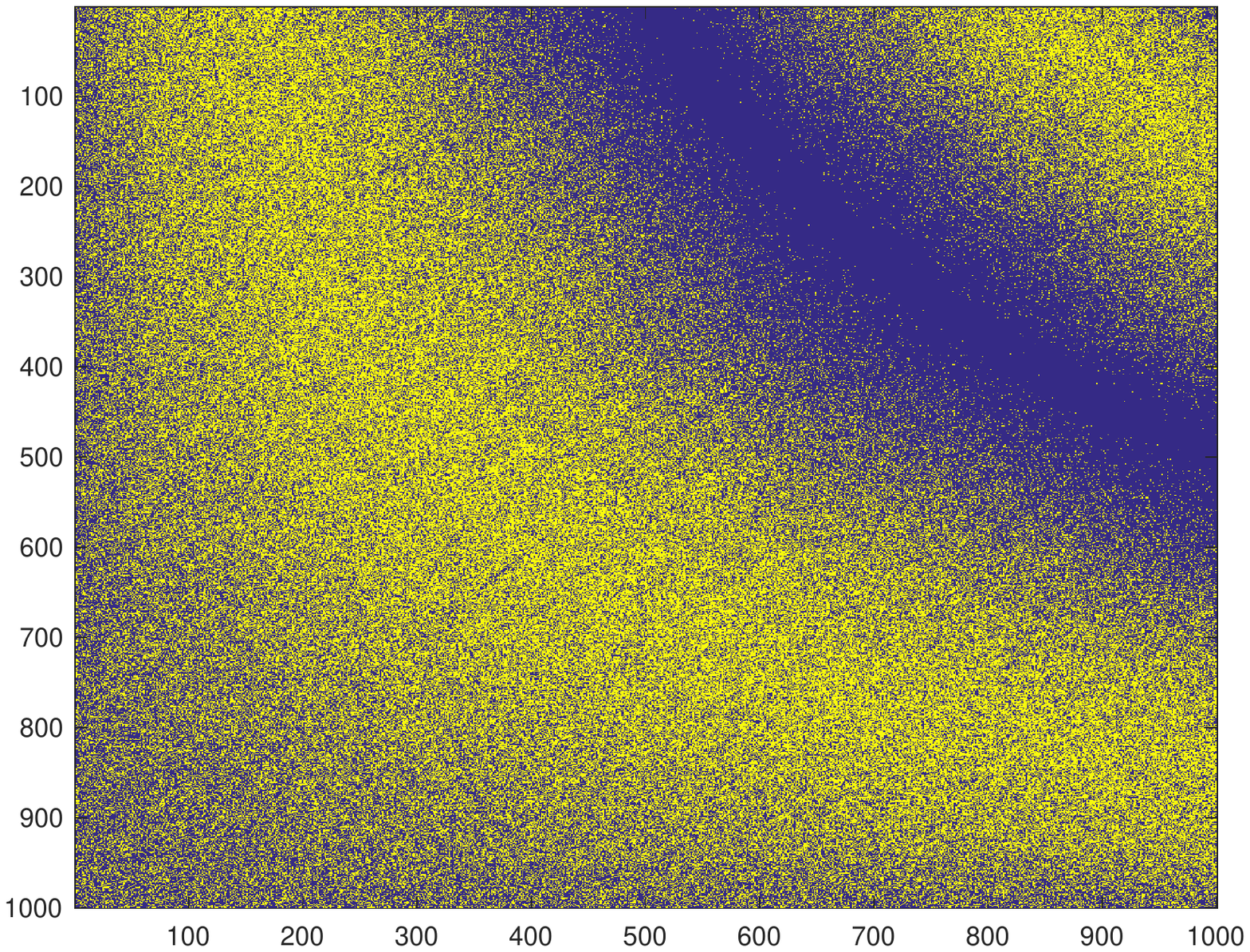}
	\end{center}
	\label{fig:sorted}
	\caption{Empirical graphon corresponding to sorted samples of $\mathcal{W}(x,y) = 1/3 + 1/3\cdot \mathrm{sin}(3\cdot \pi x \cdot y)$}.
\end{figure}

\begin{figure}[t]
	\begin{center}
		\includegraphics[width=9.3cm,trim=20 225 0 215]{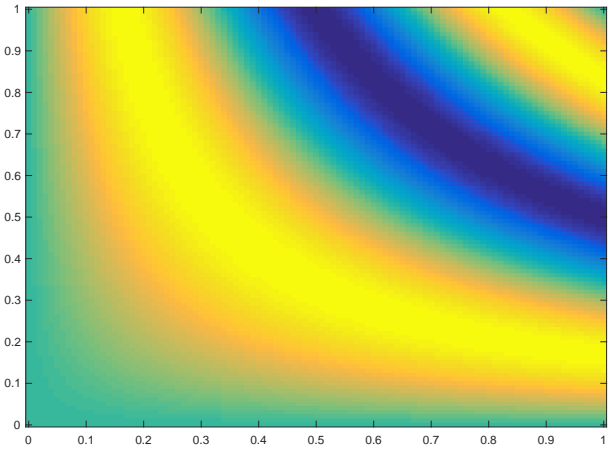}
	\end{center}
	\label{fig:graphon}
	\caption{The graphon $\mathcal{W}(x,y) = 1/3 + 1/3\cdot \mathrm{sin}(3\cdot \pi x \cdot y)$}.
\end{figure}

In Figs. $3$ and $4$ we see the effect of the sorting on $\mu_i$ in it's convergence in norm to the graphon displayed in Fig. $5$. Alternatively, we can assume that the samples $\mu_i$ might not be ordered with respect to the reals, but consistently ordered from sample to sample. In that case, the graphon to which the empirical graphon converges in norm is $\mathcal{W}_e^{\psi^*}$ where $\psi^*$ is the measure preserving map which minimizes $\|\mathcal{W} - \mathcal{W}_{e}^{\psi}\|_{\square}$.
%
Specifically, it is shown in \cite{OKNV} that the empirical graphon associated with the adjacency matrix not only converges to $\mathcal{W}(x,y)$ but is rate optimal in the minimax sense. To wit, for $k$-step graphons, it can be shown that

\begin{align}
\delta_{\square}(\mathcal{W},\mathcal{W}_e) \leq C \sqrt{\frac{k}{n \log(k)}}
\end{align}
where $C$ is a numerical constant \cite{OKNV}. In general, however, for an arbitrary graphon, the most that can be shown is a convergence rate proportional to $\log(n)^{-1/2}$ \cite{LL},\cite{OKNV}. 



What remains to be shown is that the powers $T_e^k$ converge to the power of the true graphon operator $T^k$. We proceed by proof by induction. In the following, for ease of notation we make use of the equivalence of norms to write the equations in terms of the operator 2 norm, in this case equivalent to the spectral radius.

\begin{align}
\|T^k - T_e^k\| &= \|T T^{k-1} - T_e T_e^{k-1} + T_e T^{k-1} - T_e T^{k-1}\| \nonumber \\
&= \|(T-T_e)T^{k-1} + T_e(T^{k-1} - T_e^{k-1})\| \nonumber \\
&\leq \|T^{k-1}\| \cdot \|T-T_e\| + \|T_e\|\cdot\|T^{k-1} - T_e^{k-1}\|,
\end{align}
where the final implication is a combination of Minkowski's inequality and Cauchy-Schwarz. We then proceed by induction to write

\begin{align}
\|T^{k-1} - T_e^{k-1}\| &\leq \|T^{k-2}\|\cdot\|T-T_e\| + \nonumber \\&\|T_e\|\cdot \|T^{k-2} - T_e^{k-2}\|
\end{align}

Repeating this process $k-1$ times yields the following expansion.

\begin{align}
\|T^{k} - T_e^{k}\| &\leq \|T^{k-1}\|\cdot\|T - T_e \| + \|T_e\|\cdot \|T\| \cdot \|T - T_e\| +\nonumber \\ &\cdots + \|T_e^{k-1}\|\cdot \|T - T_e\|
\end{align}
Crucially, the term $\|T - T_e\|$ is present in each summand, and the spectral radii of $T$ and $T_e$ are bounded (specifically, less than $1$). Thus, taking the limit with respect to $N$, we conclude that $\|T^k - T_e^k\| \to 0$. Defining $z(x) = (Ih_0 + \sum_{k=1}^{K}h_k T^k)f(x)$ and $z_e(x)$ as in the previous section we write

\begin{align}
\|z(x) - z_e(x)\| &= \|(\sum_{k=0}^{K} h_0 T^k - \sum_{k=0}^{K}h_0 T_e^k)f(x)\| \nonumber \\
&\leq \sum_{k=0}^{K} \|h_k\| \cdot \|T^k - T_e^k\| \cdot \|f(x)\| \to 0, \nonumber \\ &\mathrm{as}\ N\to \infty
\end{align}

In sum, $\phi$ has been shown to be bijective with $T_e f_a(x)$, which has been shown to converge to $T f(x)$. Thus, with some abuse of notation necessitated by the finiteness and infiniteness of the separate domains, we state the following theorem.
\begin{theorem}
$\|\mathbf H \mathbf x - \mathcal{H} f(x)\| \to 0$ as $N \to \infty$. $\square$ 
\label{th:conv}
\end{theorem}
where here $\mathbf H \mathbf x$ is taken to be understood as $\mathcal{H}_e f_a(x)$ where $\mathcal{H}_e = I h_0 + \sum_{k = 1}^{K} h_k T_e^k$.

\section{Shift Operator Definition}
\label{sec:FG}

In the previous sections we have demonstrated the convergence of a graph filter defined on a random graph to a graphon filter defined on the kernel-model from which the random graph was generated. In order to make the comparison between a filter defined on a finite dimensional vector space, and a filter defined on an infinite vector space, we had to introduce the concept of an operator matrix. To do this, the input and output functions and the kernel were projected onto an orthonormal basis for the space $L^2[0,1]$. Thus we arrive at the question of how to appropriately choose a basis in order to arrive at an operator which captures the behaviour of the operator $T$ as concisely as possible.

We know of a way to choose a matrix on the basis of a graphon that converges to the Fredholm integral equation: a random scaled adjacency matrix drawn from the graphon by the procedure mentioned in Section \ref{sec:random graphs} has this property. However, the convergence rate is extremely slow as we saw in the previous sections. One could choose the step-function basis used in Section \ref{sec:fredholm} however this would also require many step-functions to converge, especially for smooth graphons. This would defy our purpose stated in the introduction of how to perform graph filtering on graphs that are too large to study as a whole. Clearly, the choice of basis affects the resulting operator matrix: both in its size and complexity. Thus, the basis should be chosen such that the resulting operator matrix is as small as possible, as sparse as possible, and can be calculated as efficiently as possible.

The primary practical concern is that one should be able to solve the integrals in \eqref{eq:op_eq_trunc}, and they should be sufficiently general to be able to approximate as many different types of graphons as possible. To this end, we choose the Tchebyshev basis functions. Firstly, these functions are the best approximators of continuous functions in $L^2[0,1]$. Secondly, the Tchebyshev functions allow us to use Gauss-Tchebyshev quadrature, thus transforming the analytical integrals in \eqref{eq:op_eq_trunc} into weighted sums which can be numerically evaluated.

\subsection{Fourier-Galerkin Shift Operator}

In the Galerkin approach \cite{Delves1} to the solution of Fredholm integral equations \eqref{eq:fredholm}, a basis of orthogonal polynomials are chosen, and the function $f(x)$ and kernel $\mathcal{W}(x,y)$ are taken to be the projections onto this basis. Following \cite{Delves1},\cite{Delves2} the orthogonal basis is chosen to be the Chebyshev polynomials of the first kind, $c_i(x)$. Working under the frame work of the operator matrix vector equation \eqref{eq:op_eq}, we now obtain

\begin{align}
[\mathbf f]_i = \int_{-1}^{1} f(u) c_i(u)/(1 - u^2)^{1/2} \mathrm{d}u \label{eq:tcheb_exp}
\end{align}
and 
\begin{align}
[\mathbf W]_{i,j} = \int_{-1}^{1} \int_{-1}^{1} \mathcal{W}(u,v) c_i(u) c_j(v) / (1-u^2)^{1/2}\mathrm{d}u\ \mathrm{d}v \label{eq:tcheb}
\end{align}
Notably, the Tchebyshev functions of the first kind are only orthogonal with respect to the weight function $1/(1-x^2)^{1/2}$, hence its inclusion in equation \eqref{eq:tcheb}. However, by mapping the interval of integration from $[0,1]$ to $[-1,1]$ with the substitution rule $u = 2x - 1$ we can use the Gauss-Tchebyshev quadrature rule to convert the equations \eqref{eq:tcheb} into tractable weighted sums. Specifically, equations \eqref{eq:tcheb_exp} and \eqref{eq:tcheb} can be evaluated with the rule

\begin{align}
\int_{-1}^{1} \frac{f(u)}{\sqrt{1-u^2}} \mathrm{d}u \approx \sum_{i=1}^{N} w_i f(u_i) \label{eq:quaddef}
\end{align}
where $w_i = \frac{\pi}{N}$ and $u_i = cos(\frac{2i-1 }{N} \pi)$. Using \eqref{eq:quaddef} and using the identity

\begin{align}
c_i(u) = cos(i cos^-1 (u))
\end{align}
we can rewrite equations \eqref{eq:tcheb_exp} and \eqref{eq:tcheb} as

\begin{align}
[\mathbf f]_i = \frac{\pi}{P}\tilde{\sum_{k=1}^{P+1}} f(cos(\frac{\pi i}{N}))cos(\pi (k-1))
\end{align}

\begin{align}
&[\tilde{\mathbf{W}}]_{i,j} = \frac{\pi^2}{P^2}\tilde{\sum_{m = 1}^{P+1}} \tilde{\sum_{n = 1}^{P+1}} \mathcal{W}\bigg (\mathrm{cos}\big(\frac{\pi (m-1)}{P}\big),\mathrm{cos}\big(\frac{\pi (n-1)}{P}\big) \bigg )\nonumber \\ &\mathrm{cos}\bigg(\frac{\pi (m-1) (i-1)}{P}\bigg) \mathrm{cos}\bigg(\frac{\pi (n-1) (j-1)}{P}\bigg) \label{eq:ker}
\end{align}
where the operator $\tilde{\sum}$ denotes that the first and last terms are multiplied by $\frac{1}{2}$, and $P$ is the number of quadrature points. The operator $\tilde{\mathbf W_{i,j}}$ is, in fact, the numerical solution to the integrals

\begin{align}
\int_{-1}^{1} \frac{c_i(u)}{\sqrt{1-u^2}} \int_{-1}^{1} \frac{\mathcal{W}(u,v) c_j(v)}{\sqrt{1-v^2}} du\ dv \label{eq:tcheb_quad}
\end{align}
which has the extra weight function $\frac{1}{\sqrt{1-v^2}}$ compared to the defining equations \eqref{eq:tcheb}. Thus, approximating $T$ by \eqref{eq:tcheb_quad} will incur large errors. However, we can derive the operator $\mathbf{W}$ from the operator $\tilde{\mathbf{W}}$ by multiplying \eqref{eq:tcheb_quad} by the Tchebyshev expansion of $\sqrt{1-v^2}$

\begin{align}
\sqrt{1-v^2} &= \frac{2}{\pi} - \frac{4}{\pi} \sum_{k=1}^{\infty} \frac{c_{2k}(v)}{4k^2 -1}. \label{eq:tcheb_wt_exp}
\end{align}

By multiplying the series \eqref{eq:tcheb} by the series \eqref{eq:tcheb_wt_exp}, and using the identity

\begin{align}
c_p(u) c_q(u) = \frac{1}{2}\big(c_{p+q}(u) + c_{|p-q|}(u)\big)
\end{align}
the operator can be calculated explicitly as

\begin{align}
\frac{\pi}{2} [\mathbf{W}]_{i,j} = \tilde{[\mathbf{W}}]_{i,j} - \sum_{l=1}^{P} \frac{1}{4l^2 - 1} \big([\tilde{\mathbf{W}}]_{i,j+2l} + [\tilde{\mathbf{W}}]_{i,|j-2l|}\big) \label{eq:FG}
\end{align}
where the operator $\tilde{\mathbf{W}}$ in the above equation is equal to the operator $\tilde{\mathbf{W}}$ in \eqref{eq:ker} but padded with extra zeros depending on the length of the series \eqref{eq:tcheb_wt_exp}. The calculation of the Fourier-Galerkin shift operator is summarized in Algorithm $1$.

\begin{algorithm}
	\label{alg:Fourier Galerkin}
	\begin{algorithmic}
		\State calculate $\tilde{\mathbf W}$ with \eqref{eq:ker}
		\State calculate $\mathbf W$ with \eqref{eq:FG}
		\State return $N \times N$ principal submatrix of $\mathbf W$
	\end{algorithmic}
	\caption{Fourier-Galerkin Shift Operator ($\mathcal{W}(x,y)$,$P$,$N$)}
\end{algorithm}

\textit{Illustrative Example}: The simplest example, structurally speaking, is that of the Erd\"os-Reny\'i (ER) model. Applying \eqref{eq:tcheb} to the function $\mathcal{W}(x,y) = p$ readily gives the result that $[\mathbf{W}]_{1,1} = \pi^2 p$, with $[\mathbf W]_{i,j} = 0,\ \forall i \neq j$ and $i,j \geq 1$, since $\sum_{m=1}^{M} \mathrm{cos}(\frac{\pi (m-1) (i-1)}{P}) = 0$, with the exception of $i = 1$. Therefore, diffusion over an Erd\"os-Reny\'i graph, as captured by multiplication by the scaled adjacency matrix, can be encoded with a single coefficient. Mathematically, the point becomes clear by investigating the resolvent equation corresponding to an ER kernel

\begin{align}
\lambda f(x) = \int_{0}^{1} p \boldsymbol{1}(x,y) f(y)\ \mathrm{d}y
\end{align}
the solutions to this equation are clearly $\lambda = p$, which corresponds to the eigenfunction $\boldsymbol{1}(x)$, and $\lambda = 0$ with infinite multiplicity, which corresponds to any function which integrates to $0$. 

\begin{figure}[t]
	\begin{center}
		\includegraphics[width=9.3cm,trim=20 225 0 215]{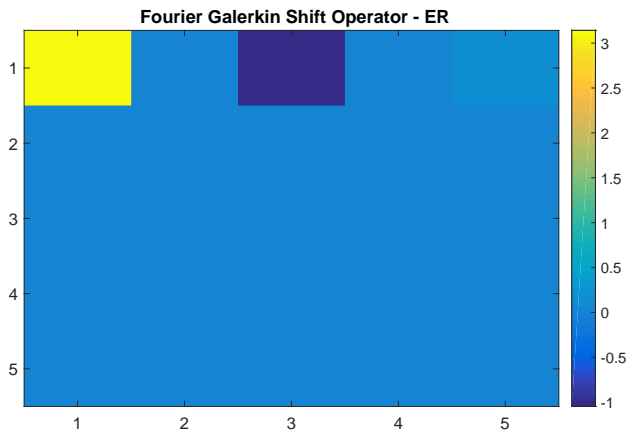}
	\end{center}
	\caption{Fourier-Galerkin shift operator corresponding to an Erd\"os-Reny\'i model with $p = 0.5$. There is one non-zero row, the first, indicating that the range space of this operator is the constant functions.}
	\label{fig:FGSO}
\end{figure}

\begin{figure}[t]
	\begin{center}
		\includegraphics[width=9.3cm,trim=20 225 0 215]{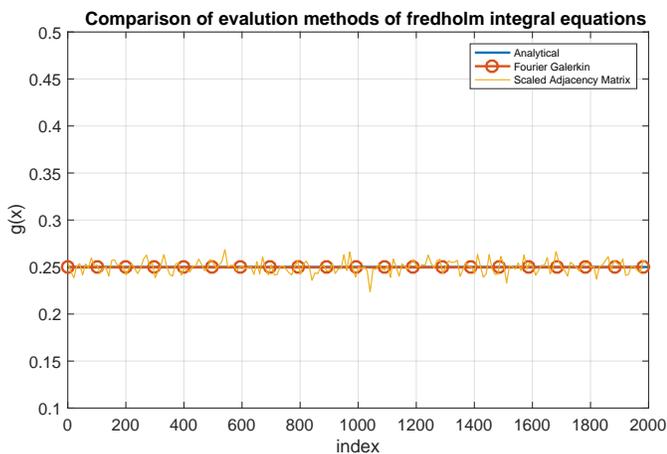}
	\end{center}
	\caption{Evaluation of the integral equation $\int_{0}^{1} p \boldsymbol{1}(x,y) y \mathrm{d}y$ for $p = 0.5$ via analytical solution, Fourier-Galerkin method, and diffusion through the scaled adjacency matrix of an ER graph with parameter $p = 0.5$ and $N = 2000$ nodes}.
	\label{fig:FG_SO_ER}
\end{figure}

Fig. \ref{fig:FGSO} shows the Fourier-Galerkin shift operator corresponding to an ER random graph model with parameter $p = 0.5$. A total of 5 Tchebyshev functions and 10 quadrature points were used in the evaluation of \eqref{eq:tcheb_quad}. It can be observed that the matrix $\mathbf W$ has only one non-zero row, specifically the first row, indicating that the range space of the operator is the set of constant functions. To demonstrate that the matrix $\mathbf W$ is the correct linear map corresponding to the kernel function $p = 0.5$ we first must assume a function $f(y)$. We thus assume the function $f(y) = y$ and evaluate the integral $\int_{0}^{1} p \boldsymbol{1}(x,y) y \mathrm{d}y$ having the obvious solution $\frac{1}{4}\boldsymbol{1}(x)$. A close correspondence between the analytical solution, the Fourier-Galerkin solution, and the evaluation via the scaled adjacency matrix is observed in Fig. \ref{fig:FG_SO_ER}.

\begin{figure}[t]
	\begin{center}
		\includegraphics[width=9.3cm,trim=20 225 0 215]{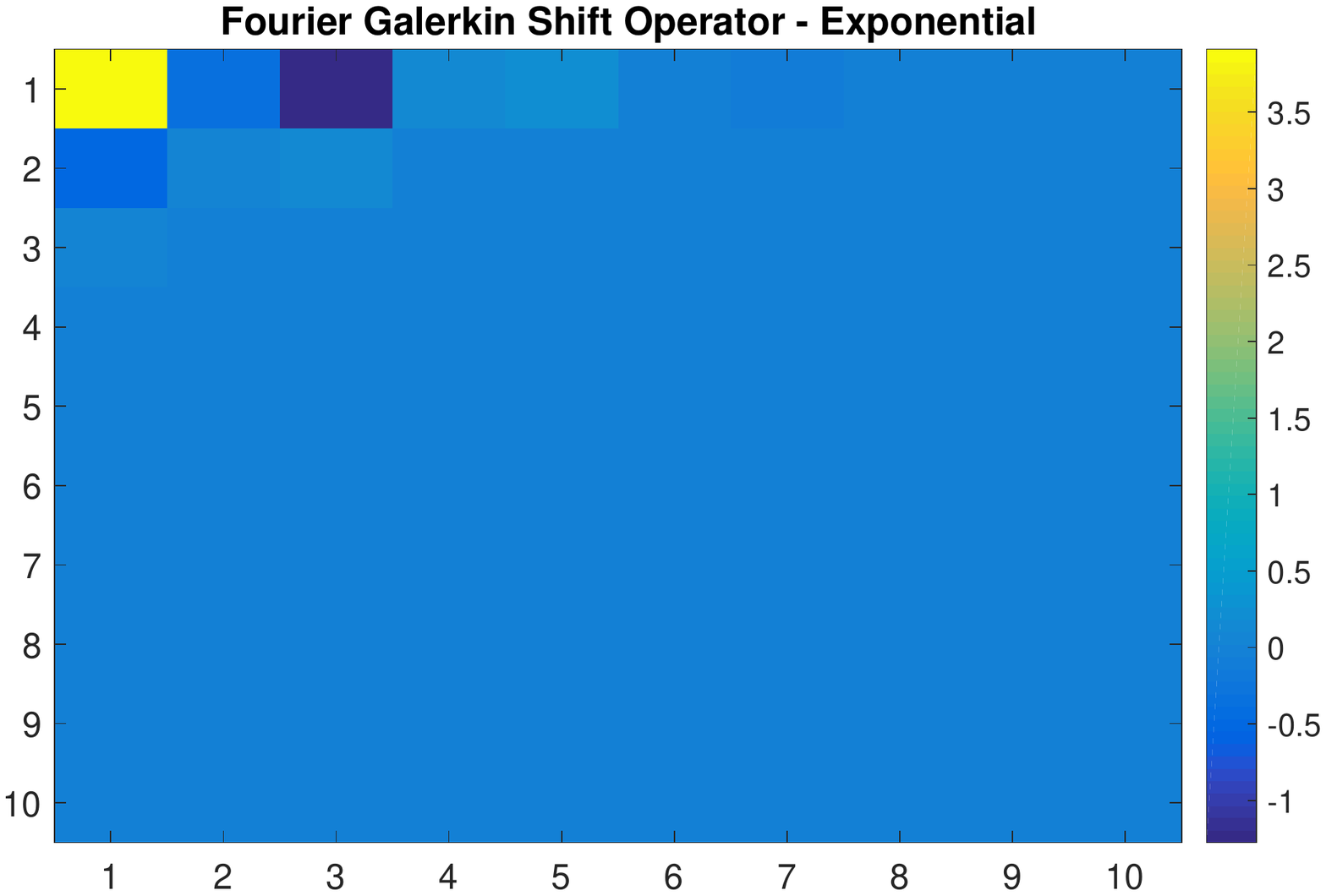}
	\end{center}
	\label{fig:FG_exp}
	\caption{Fourier-Galerkin shift operator corresponding to the graphon $e^{(-1/2(x+y))}$. There are non-zero coefficients in more than the first row.}
	\vspace{0.6cm}
	\begin{center}
		\includegraphics[width=9.3cm,trim=20 225 0 215]{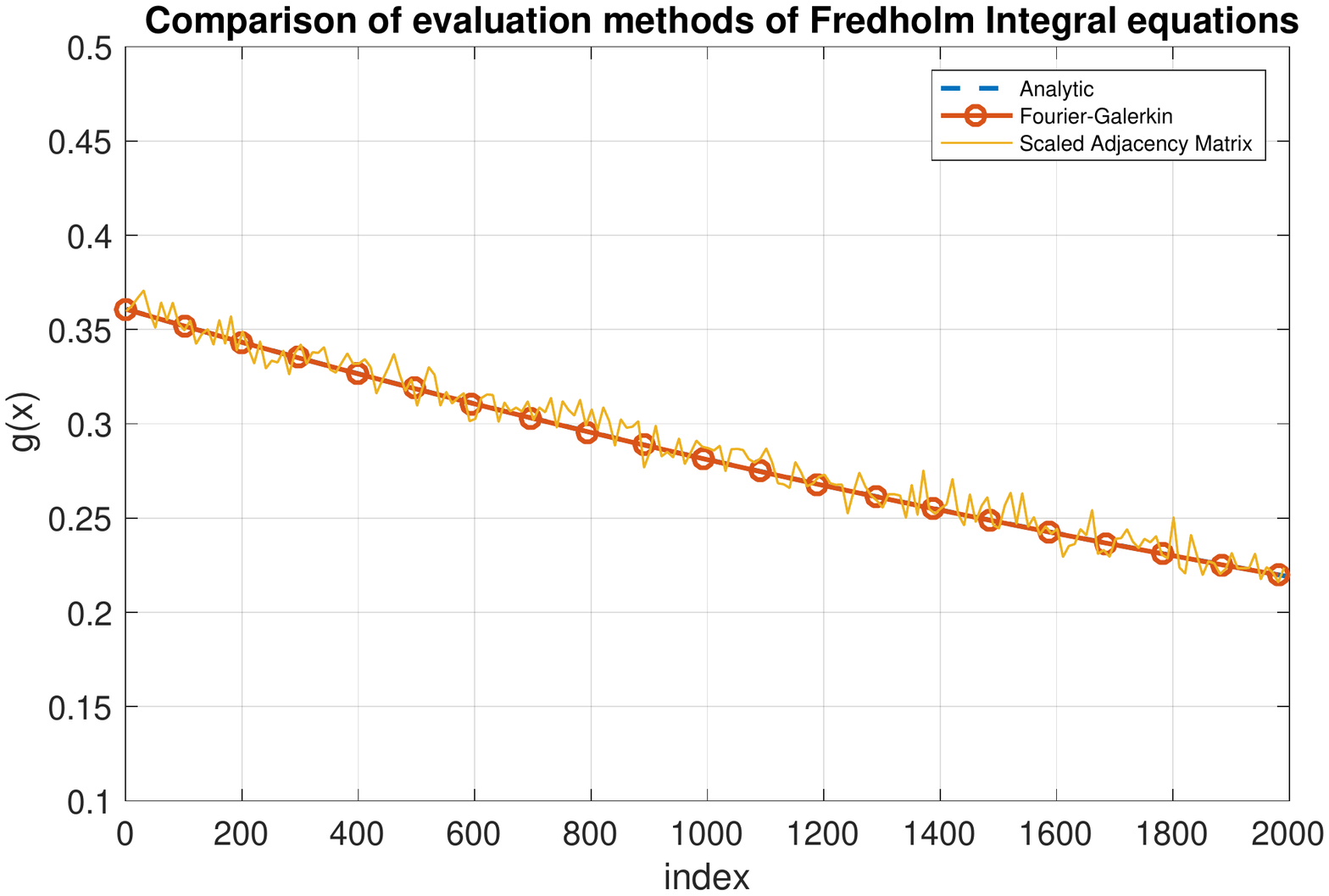}
	\end{center}
	\label{fig:FGSO_exp}
	\caption{Evaluation of the integral equation $\int_{0}^{1} e^{(-1/2(x+y))} y \mathrm{d}y$ via analytical solution, Fourier-Galerkin method, and diffusion through the scaled adjacency matrix with $N = 2000$ nodes.}
\end{figure}

Figs. \ref{fig:FG_exp} and \ref{fig:FGSO_exp} show the Fourier-Galerkin shift operator corresponding to an exponential random graph model with kernel function $e^{-\frac{1}{2}(x+y)}$ and the evaluation of the integral equation $\int_{0}^{1} e^{-\frac{1}{2}(x+y)} y \mathrm{d}y$ which has the solution $(4 - \frac{6}{e^{1/2}}) e^{-1/2x}$, respectively. A close correspondence between the evaluation of the integral equation by the scaled adjacency matrix, Fourier-Galerkin method, and the analytical solution is again observed.

\begin{myremark}The convergence of the scaled adjacency matrix to its limit object occurs as the number of nodes and connections in the network grows. The convergence of the solution of the integral equation by quadrature to the analytical solution only depends on the complexity of the kernel. Thus, given an arbitrarily large graph, given the Kernel function explaining the graph, the Fourier-Galerkin shift operator achieves a compression of constant order (with respect to the graph size) of the random graph. Additionally, given the properties of approximation by Tchebyshev functions, the approximation error of the solution of the Fredholm integral equation can be made arbitrarily small. The error in approximation of the graph filter by the graphon analog is thus dominated by the convergence of the scaled adjacency matrix to the limit object, the Fredholm integral equation.\end{myremark}

\section{Graphon Filter Design}
\label{sec:GFD}

In this section, we address the design of simple graphon filters given the graphon. As was shown in the previous section, the Fourier-Galerkin shift operator is a linear map between the Tchebyshev representation of the input function to a Tchebyshev representation of the output function (the solution of the integral equation \eqref{eq:fredholm}). In essence, filtering in this context is thus controlling what Tchebyshev representations are possible as an output of the evaluation of the filter operator. As such, the goal of designing a graphon filter is the attenuation of one or more of the Tchebyshev coefficients of the output.

The first step of the graphon filtering algorithm is to create a representation of the input graph signal $\mathbf x$. To do this, we set $f(x) = \sum_{i=1}^{N} [\mathbf x]_i b_i'(x)$, where $b_i'(x)$ is the step basis functions used in Section \ref{sec:fredholm} supported on $[-1,1]$. Then, to produce $\mathbf f$ we use equation \eqref{eq:tcheb_exp}. 

The graph ``frequencies'' correspond to the various Tchebyshev functions respectively. Thus, an ``all pass'' graphon filter is simply the identity matrix. A ``frequency'' band is analogous to selecting a certain number of rows of the identity matrix. Thus, the ideal graphon filter response is contained in the diagonal of a matrix $\mathbf D$. Using the graphon filter definition proposed in section \ref{sec:HSop} in combination with the convergence of the scaled adjacency matrix to a Fredholm integral equation investigated in the previous section, a simple graphon filter $\mathbf H$ is a polynomial in the Fourier-Galerkin shift operator.

\begin{align}
\mathbf H = \sum_{k = 0}^{K-1} h_k \mathbf W^k
\end{align}
The design of the simple graph filter can be stated as a simple least squares problem

\begin{align}
\min_{\{h_k\}},\ \ 
\|\sum_{k=0}^{K-1} h_k  \mathbf W^{k} - \mathbf{D} \|_F^2 \label{eq:graphon_filter_def}
\end{align}
which can be restated in matrix vector form via the vectorization operator as
\begin{align}
\min_{\mathbf h}, \ 
\|\mathbf A \mathbf h - \mathbf b\|_2^2 \nonumber
\end{align}
where the $k$-th column of $\mathbf A$ is $\mathrm{vec}(\mathbf W^{k-1})$ and $\mathbf b = \mathrm{vec}(\mathbf D)$, which is solved optimally via least squares. Due to the polynomial nature of the columns of $\mathbf A$, the matrix $\mathbf A$ will be highly ill-conditioned resulting in inaccurate coefficients of extreme magnitude. Thus, for reasons of numerical stability, the graphon filter should be designed via truncated SVD.

\begin{algorithm}
	\begin{algorithmic}
		\State Produce $\mathbf f$ with \eqref{eq:tcheb_exp}
		\State $\mathbf A = [\mathrm{vec}(\mathbf W^0) \mathrm{vec}(\mathbf W) \cdots \mathrm{vec}(\mathbf W^{k-1})],\ \mathbf b = \mathrm{vec}(\mathbf D)$
		\State $\min_{\mathbf h} \|\mathbf A \mathbf h - \mathbf b\|$
		\State return $\mathbf h = \mathbf A^{\dagger} \mathbf b$
		\State $\mathbf g = \mathbf H \mathbf f$
		\State $\mathbf y' = \sum_{n=1}^{N} [\mathbf{g}]_n \mathbf{t}_n$
	\end{algorithmic}
	\caption{Graphon Filter ($\mathbf W$,order $k$,$\mathbf D$)} 
	\label{alg:graphon_filter}
\end{algorithm}

To move back from the graphon ``frequency'' domain to the graph node domain, we use resampling. That is, given the Tchebyshev representation of the input $f$, and the implementation of the filter $\mathbf g = \mathbf H \mathbf f$, the node domain representation would be given by the $T$ dimensional vector

\begin{align}
\mathbf y' = \sum_{n=1}^{N} [\mathbf{g}]_n \mathbf{t}_n
\end{align}
where $\mathbf t_n$ is $n$-th first-kind Tchebyshev function sampled at $T$ points uniformly over $[-1,1]$, and $\mathbf y'$ denotes the resampled node domain graphon filtered output, as opposed to the graph filtered output $\mathbf y = \sum_{k=0}^{K-1} h_k \mathbf S^k \mathbf x$. Crucially, we point out that the coefficients designed by the graphon filtering algorithm are the coefficients for the graph filtering operation. These steps are summarized in Algorithm \ref{alg:graphon_filter}.

\subsection{Filter Response}

One interpretative benefit of the graphon filter algorithm is that filtering is accomplished in terms of a standard basis. Specifically, while different graphons (and therefore, the random graphs drawn from the graphons) may have different representations in this standard basis, but the output will always be in the standard basis. In this sense, a ``low pass'' or ``high pass'' filter has the same meaning regardless of the graph structure. The output of the filter will only have non-zero coefficients corresponding to the desired basis functions. This does not have a parallel to graph filtering using the eigenbasis of a graph shift operator, wherein the graph filtering operation and output are graph dependent. In contrast, the graphon filtering algorithm is not only agnostic with respect to the particular graph, but also with respect to the graphon. Only the reachability of $\mathbf D$ depends on the graphon. 

Additionally, the matrix operator provides a clear interpretation of the frequency response of the graphon filter. Since the amplitudes of the input frequencies are the $j$-th coefficient of the input vector $\mathbf f$, the frequency response of the graphon filter $\mathbf H$ is the output of $\mathbf H \boldsymbol{1}$. This can easily be seen by considering the response to each frequency individually. The response of $\mathbf H$ to the $j$-th frequency is obviously $\mathbf H \mathbf e_j$, and the response to all frequencies is just the sum over all frequencies $j$, which is exactly $\mathbf H \boldsymbol{1}$. In contrast to classical signal processing, the frequencies are discrete. There are no graph frequencies ``between'' the Tchebyshev frequencies. Thus the frequency response is not a continuous function, but a vector. 

\begin{figure}[t]
	\begin{center}
		\includegraphics[width=9.3cm,trim=20 225 0 215]{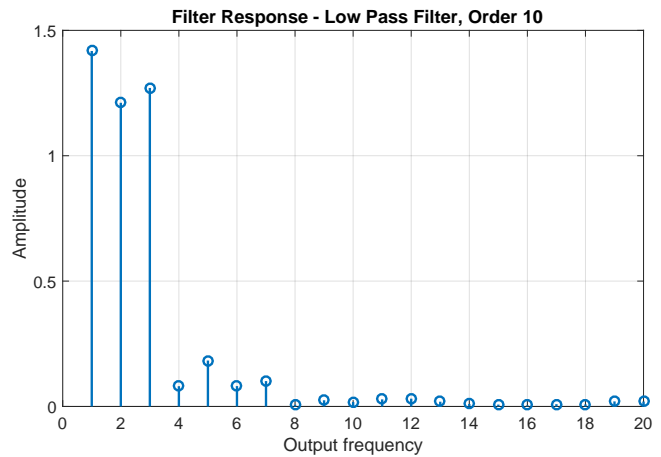}
	\end{center}
	\caption{Low-pass filter response of a 10th order graphon filter implemented over a kernel of $e^{-5|x-y|}$. The filter was designed to preserve the first three graphon frequencies while nullifying any higher frequencies.  }
	\label{fig:filt_resp1}
\end{figure}

\begin{figure}[t]
	\begin{center}
		\includegraphics[width=9.3cm,trim=20 225 0 215]{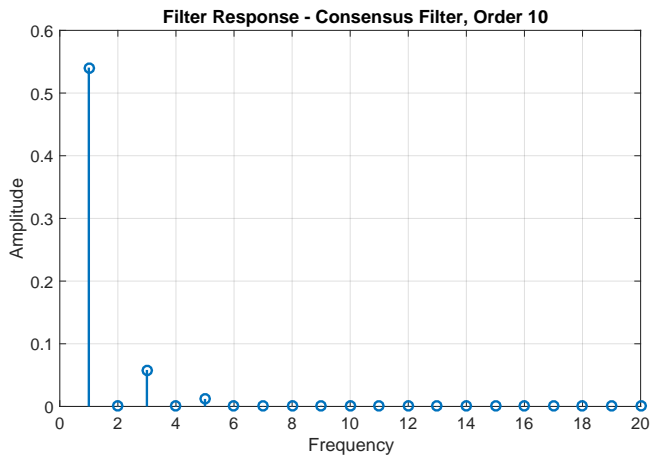}
	\end{center}
	\caption{Consensus filter response of a 10th order graphon filter implemented over a kernel of $e^{-5|x-y|}$. The filter was designed to have as small of an output as possible for all except the constant frequency.  }
	\label{fig:filt_resp2}
\end{figure}

Fig. \ref{fig:filt_resp1} shows the frequency response of a ``low-pass'' graphon-filter designed for the graphon $e^{-5|x-y|}$. Fig. \ref{fig:filt_resp2} shows the frequency response corresponding to a ``consensus'' filter designed for the same graphon. Comparing the two figures we can clearly see that the graphon filtering algorithm is able to both preserve and nullify the selected frequencies and thus implement a variety of filtering tasks.

\section{Simulations}
\label{sec:sims}

In order to demonstrate the graphon filtering algorithm, the effects of the graphon on the design and reachability of certain filters, as well as the convergence results presented in this paper, we examine three graphons: and ER-graphon with $p=0.5$, $\frac{1}{2} + \frac{1}{2}\mathrm{sin}(\frac{7}{2}\pi x y)$, and $e^{-10|x-y|}$. Two separate filtering tasks are implemented: ``low-pass'' filtering, wherein the aim is to preserve several low frequencies, and consensus filtering where the aim is to preserve only the constant frequency present in the graph signal. All graph filters are implemented with the coefficients designed by Algorithm $2$, without knowledge of the specific graph sample. The input function is $f(x) = x+sin(x)$. Empirical samples are drawn with $N = 2000$ nodes.

\subsection{Low Pass Filter}

\begin{figure}[t]
	\begin{center}
		\includegraphics[width=9.3cm,trim=20 225 0 215]{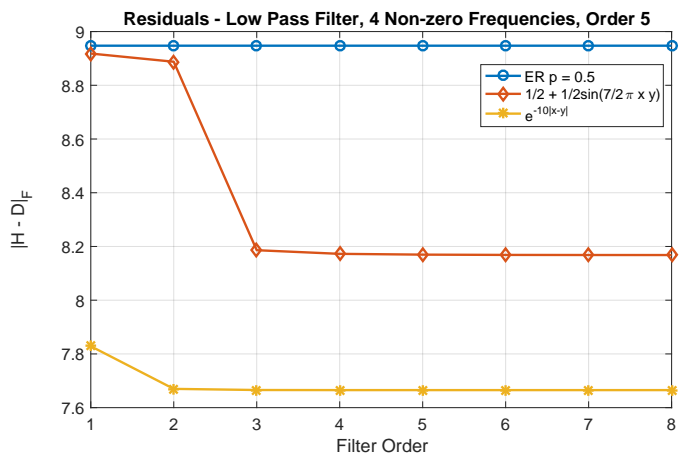}
	\end{center}
	\caption{Residuals of the least squares filter design algorithm for a Low Pass filter. A large difference in the ability of different random graph models to implement the ``low pass'' filtering task is observed. }
	\label{fig:LowPassRes}
\end{figure}

We implement the graphon filtering algorithm described in Algorithm $2$ for the aforementioned three graphons, for orders $1$ through $8$, in order to implement a low-pass graphon filter. The matrix $\mathbf D$ in the least squares formulation in Algorithm $2$ was here specified to be $\mathrm{diag}(\mathbf b)$ where $\mathbf b = [1,5,5,10,0,\cdots,0]$. Fig. \ref{fig:LowPassRes} shows the residuals from the least squares filter design algorithm.

Inspecting Fig. \ref{fig:LowPassRes} it becomes clear that the underlying graphon has an enormous effect on the ability to match the desired graphon filter operator. With an ER graphon, Algorithm $2$ is completely unable to match the desired graphon filter operator $\mathbf D$. This is because the related Fourier-Galerkin shift operator has no higher order frequency content. Thus all higher order frequencies are filtered out by default, leaving only the constant frequency. Other graphon models allow Algorithm $2$ to match the ideal graphon filter operator, specifically because these models have higher frequency content. The graphon with the highest frequency content here considered, $e^{-10|x-y|}$ is the best able to match the ideal graphon filter. Moreover, the graphon with the higher frequency content is able to implement its filter design with a lower order filter.

\begin{figure}[t]
	\begin{center}
		\includegraphics[width=9.3cm,trim=20 225 0 215]{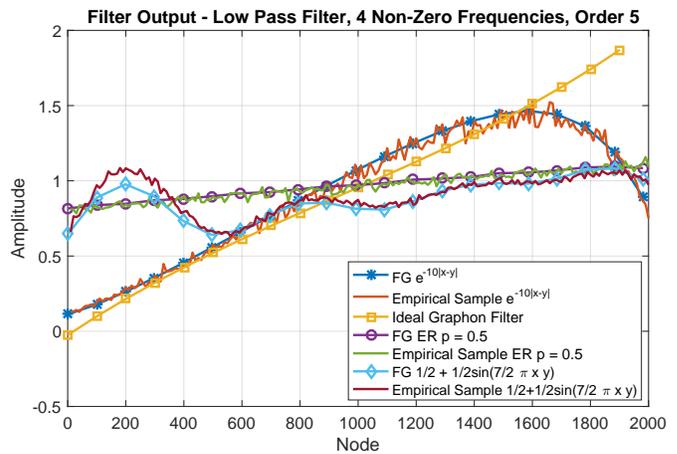}
	\end{center}
	
	\caption{The output of 3 order 5 graphon filters. The first 4 graphon frequencies were to be preserved while the rest were to be nullified. The input function was $y + sin(y)$ over the domain $[0,1]$.} 
	\label{fig:LowPassOutput}
\end{figure}

Fig. \ref{fig:LowPassOutput} shows the output of low pass filters designed over the aforementioned three graphons, with order $5$. As Theorem \ref{th:conv} predicts, the output of all graphon filters closely track the empirical samples. The output of the ideal filter is plotted for comparison purposes. It can be observed that the output of the graphon filter designed over $e^{-10|x-y|}$ fits closest to the output of the ideal filter. Whereas the other two filters have much lower frequency content they have much more difficulty in matching the ideal filter output.

\subsection{Consensus Filtering}

\begin{figure}[t]
	\begin{center}
		\includegraphics[width=9.3cm,trim=20 225 0 215]{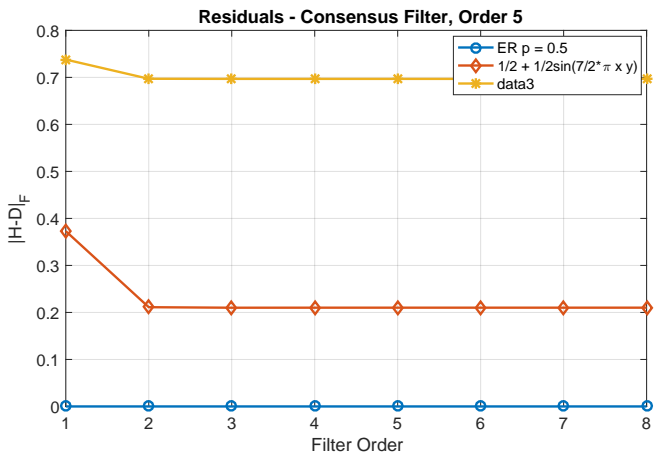}
	\end{center}
	\caption{Residuals of the least squares filter design algorithm for a consensus filter. A large difference in the ability of different random graph models to implement the ``consensus'' filtering task is observed. }
	\label{fig:ConsFiltRes}
\end{figure}

We again implement the graphon filtering algorithm described in Algorithm $2$ but this time in order to implement a consensus filtering task. The Algorithm is again implemented over the same $3$ graphons. The ideal graphon filter for this operation is $\mathbf D = \mathrm{diag}(\mathbf e_1)$. Fig. \ref{fig:ConsFiltRes} shows the residuals from the least squares consensus filter design algorithm.

Contrary to what was observed in the low pass example, the graphon filtering algorithm is able to almost-perfectly implement the consensus filter. In direct contrast to the low pass filtering task, higher frequency content in the Fourier-Galerkin shift operator is undesirable as these higher order frequencies will need to be nullified. In this case the higher order frequencies cannot be effectively nullified using a simple graphon filter. 

\begin{figure}[t]
	\begin{center}
		\includegraphics[width=9.3cm,trim=20 225 0 215]{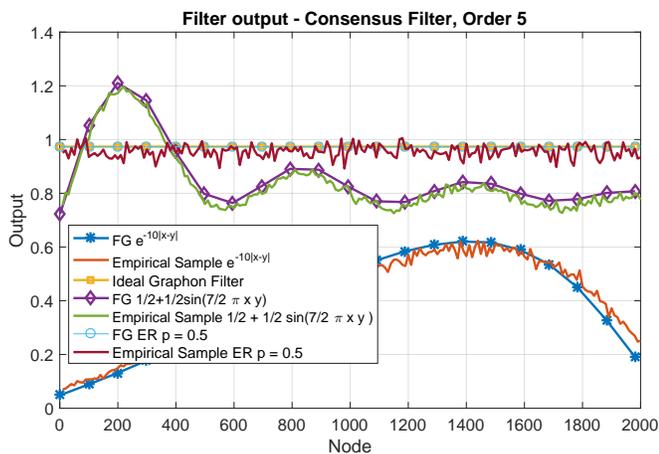}
	\end{center}
	\caption{The output of three order $5$ graphon consensus filters. The input function was $y + sin(y)$ over the domain $[0,1]$.  }
	\label{fig:ConsFiltOut}
\end{figure}

Fig \ref{fig:ConsFiltOut} shows the output of graphon consensus filters designed over the aforementioned graphons compared with the output of the ideal graphon filter. Again we see a close correspondence between the empirical graph filter and the prediction made by the graphon filter. Contrary to the previous example, random graphs drawn from the graphon with the highest frequency content, $e^{-10|x-y|}$, have the most trouble performing the consensus task, while the graphon with the least frequency content, the Erd\"os-Reny\'i, performs the consensus task with a high degree of accuracy. The excess frequency content presents a challenge for the simple graphon filtering procedure. In order to overcome the structural challenges presented by the various graphon filtering tasks, the action of the nodes must be altered in some way: either via a node or edge varying filter, or a different shift action, or both.

\section{Conclusion}

In this report we defined several concepts. Firstly, we defined the concept of a graphon filter, both in discrete and continuous time. Secondly, the sparse approximation of Fredholm integral equations allowed us to define the relationship between diffusion of a signal on a graph, represented by the graph shift operator, and a Fredholm integral equation with the empirical graphon as its kernel. The approximation error was shown to be a function of an integral quadrature error, and the spectral radius of the empirical graphon. Finally, it was shown that the solution of the Fredholm integral equation defined by the empirical graphon converges to the solution of the Fredholm integral equation defined by the continuous graphon. Simulation results verified these theoretical results.

\bibliographystyle{IEEEbib}

\end{document}